\documentclass[letterpaper,twocolumn,10pt]{article}
\usepackage{usenix2019_v3}

\usepackage[colorinlistoftodos]{todonotes}
\usepackage[utf8]{inputenc}
\usepackage{cite}
\usepackage{algorithmic}
\usepackage{cleveref}
\usepackage{algorithm2e}
\usepackage{graphicx}
\usepackage{siunitx}
\usepackage{multirow}
\usepackage{textcomp}
\usepackage{xcolor, colortbl}
\usepackage{subcaption}
\usepackage{amsmath,amssymb,amsfonts}
\usepackage{amsthm}
\newtheorem*{def_adv_ex}{Definition: Adversarial Examples}

\begin{document}

\date{}

\title{\Large \bf DLA: Dense-Layer-Analysis for Adversarial Example Detection}

\author{
    Philip Sperl, Ching-Yu Kao, Peng Chen, Konstantin Böttinger\\\
   Fraunhofer Institute for Applied and Integrated Security
    \\\{philip.sperl, ching-yu.kao, peng.chen, konstantin.boettinger\}@aisec.fraunhofer.de
}

\maketitle

\begin{abstract}
In recent years Deep Neural Networks (DNNs) have achieved remarkable results and even showed super-human capabilities in a broad range of domains.
This led people to trust in DNNs' classifications and resulting actions even in security-sensitive environments like autonomous driving.

Despite their impressive achievements, DNNs are known to be vulnerable to adversarial examples.
Such inputs contain small perturbations to intentionally fool the attacked model.

In this paper, we present a novel end-to-end framework to detect such attacks during classification without influencing the target model's performance.
Inspired by recent research in neuron-coverage guided testing we show that dense layers of DNNs carry security-sensitive information.
With a secondary DNN we analyze the activation patterns of the dense layers during classification runtime, which enables effective and real-time detection of adversarial examples. 

Our prototype implementation successfully detects adversarial examples in image, natural language, and audio processing. 
Thereby, we cover a variety of target DNNs, including Long Short Term Memory (LSTM) architectures. 
In addition, to effectively defend against state-of-the-art attacks, our approach generalizes between different sets of adversarial examples. 
Thus, our method most likely enables us to detect even future, yet unknown attacks.
Finally, during white-box adaptive attacks, we show our method cannot be easily bypassed.
\end{abstract}

\section{Introduction}\label{sec:introduction}
Machine learning (ML) and especially deep learning (DL) applications transform modern technologies at an impressive pace.
Research progress and the availability of high performance hardware enable the training of increasingly complex models.
Such DL models have achieved even super-human results in a broad range of domains:
From classical image classification tasks \cite{DBLP:conf/cvpr/SzegedyVISW16}, to outplaying humans in Go \cite{44806}, or even autonomously driving cars \cite{DBLP:journals/corr/BojarskiTDFFGJM16}.

In numerous scenarios the security and safety are of crucial importance.
Errors in the ML processing pipeline can affect our daily routine, lead to severe incidents in the users' health, or threaten future critical infrastructures.
Such errors not only stem from inaccuracies in the training phase, but also from intentionally performed attacks.
Kurakin et al. \cite{45816} for example showed the vulnerability of self driving cars and demonstrated a successful attack.
Hence, the security of systems incorporating DL concepts is a major task for engineers, data scientists, and the research community.

Malicious actions aiming at DL models come in two flavours according to their attack timing. 
\textit{Poisoning} attacks target the training phase, while \textit{evasion} attacks are performed in the test phase.
For \textit{poisoning} attacks the attacker induces changes to the training dataset and especially to the labels to provoke misclassifications, see \cite{6868201,Nelson:2008:EML:1387709.1387716}.
As the training dataset is typically not available for attackers, the majority of recent work focuses on \textit{evasion} attacks.
Here the attacker manipulates the behavior of the DL model itself such that intended misclassifications occur.
In 2014, Szegedy et al. \cite{42503} first demonstrated that small perturbations on images fed to a deep neural network (DNN) can provoke such a misclassification.
Since then, new attacks and countermeasures have been introduced at a fast pace without the discovery of a fundamental and general defense strategy, yet.
Perturbed inputs which successfully fool the target network are known as \textit{adversarial examples}.
In  this paper, we propose an effective defense mechanism that detects such adversarial example attacks with high accuracy. 
Our approach generalizes between a broad range of state-of-the-art attacks and therefore does not only cover contemporary attacks, but will most likely also defend against future attacks. 
Further, our method defends against attacks in image classification, natural language, and even audio processing scenarios.

Currently, adversarial attacks seem to subdue corresponding defence methods.
Research in this field is yet to provide a generally applicable solution to this problem, which motivates the work in this paper.
Our main idea is based on observing neural activity during classification runtime. 
We were inspired by recent findings in the field of neural network testing and its interesting prospects.
Pei et al. \cite{DBLP:journals/corr/PeiCYJ17} introduced the idea of neuron coverage, which serves as a metric to guide testing of neural networks.
Since then, further coverage metrics have been proposed and various testing techniques have made use of them \cite{DBLP:journals/corr/abs-1803-07519, DBLP:journals/corr/abs-1803-04792}. 
Odena and Goodfellow \cite{tensorfuzz} reported promising results when applying concepts of coverage-guided fuzzing to neural network testing using neuron coverage.

These recent findings indicate that the neuron coverage of DL models carry security-sensitive information.
This hypothesis at hand led us to the main insight of this paper: 
we show that neuron coverage exhibits a characteristic behavior when processing adversarial examples. 
In particular, adversarial examples provoke a unique pattern in the coverage such that respective inputs become detectable. 
Interestingly, this characteristic is independent of the attack method, as our results strongly indicate.
With this observation we optimistically assume that our approach will also defend against future unknown attacks.

In summary we make the following contributions:
\begin{itemize} \setlength\itemsep{-0.2em}
    \item We propose a general end-to-end method to detect adversarial examples generated using different state-of-the-art methods.
    \item We successfully detect adversarial examples in image classification, natural language processing (NLP) and DL-based audio processing.
    \item We implement and evaluate our approach to successfully detect prior unseen adversarial examples of various attack methods.
    \item Finally, we evaluate our method during adaptive attacks and achieve superior results compared to related methods.
\end{itemize}

The rest of this paper is organized as follows. 
In Section \ref{related_work}, we review related work and summarize latest findings on defense strategies against adversarial attacks.
We present our main contribution, a novel concept of detecting adversarial attacks on neural networks, in Section \ref{Methodology}.
Sections \ref{Experiments} and \ref{Results} present a thorough proof-of-concept including experiments and evaluation of the according results.
For future work on this topic, we plan to publish our code and used datasets.
In Section \ref{Discussion}, we discuss the cost of our method as well as the add-on analysis regarding the real-world application, transferability, and generalisation to future attacks. 
Finally, we conclude the paper with Section \ref{Conclusion}.

\section{Related Work and Background} \label{related_work}
In this section, we discuss related work and the theoretical background that forms the basis of our approach. 
We start with adversarial attacks and discuss corresponding state-of-the-art defense methods.

\subsection{Adversarial Attacks} \label{adversarial_attacks}
We focus on test-time attacks exclusively and therefore define the following categories as introduced in \cite{DBLP:journals/corr/abs-1810-00069}.
The different evasion attack types differ in the amount and nature of information available for the attacker.
In \textit{white-box} attacks the attacker has full control over the target which includes knowledge about the architecture and parameters of the trained model.
Hence, the adversary is able to deliberately craft adversarial examples exploiting the knowledge of the model.
Contrary to that, in \textit{black-box} attacks, the attacker has neither knowledge of the target model architecture nor access to the parameters after training.
In the following, state-of-the-art attacks belong to the class of \textit{white-box} methods.
The aim of an evasion attack is to generate an adversarial example that is misclassified by the targeted DL model. 
More formally:
\begin{def_adv_ex}
Let f($\cdot$) be a trained neural network used for classification tasks. 
Let H($\cdot$) be a human oracle with the same classification capabilities.
Assume that for a given legitimate input $x$ the following equation holds:
\begin{equation*}
f(x) = H(x)
\label{eq1}
\end{equation*}
Let $x'$ be a mutated version of $x$ that is close to $x$, i.e., $\|x'-x\| \leqslant \epsilon$ for some small $\epsilon\in\mathbb{R^+}$. 
Then $x'$ is an adversarial example, if the following holds:
\begin{align*}
H(x) = H(x')\  \  \wedge \  f(x') \neq H(x').
\end{align*}
\label{eq2}
\end{def_adv_ex} 
Informally, adversarial examples are slightly mutated versions of their original counterparts that lead the targeted network to misclassification.

Szegedy et al. \cite{42503} first demonstrated the vulnerability of neural networks to slightly mutated inputs.
The authors formulated the problem of finding those mutations with a minimization problem.
To solve this problem the authors used a box-constrained \textit{L-BFGS} \cite{Flet87}.

In 2014, Goodfellow et al. \cite{43405} picked up the previous findings and proposed their resulting ``Fast Gradient Sign Method'' (\textit{FGSM}). 

Kurakin et al. \cite{DBLP:journals/corr/KurakinGB16} proposed the ``Basic Iterative Method'' (\textit{BIM}). 
In this attack, the inputs are mutated based on single steps which aim to increase the loss function.
After each step the direction is adjusted.

Madry et al. \cite{DBLP:conf/iclr/MadryMSTV18} further refined the approach.
The authors showed the BIM attack being equivalent to ``Projected Gradient Descent'' (\textit{PGD}).
By making use of the $L_{\infty}$ version of this standard convex optimization method the authors further improved the previously shown BIM.

Moosavi-Dezfooli et al. \cite{DBLP:journals/corr/Moosavi-Dezfooli15} proposed \textit{DeepFool}, which generates adversarial perturbations by iteratively pushing the inputs towards the decision boundary of the attacked network.
In order to model the decision boundary in a simplified manner, it is linearized and represented using a polyhydron.

The majority of current attacks are restricted by the $L_{\infty}$ or $L_{2}$ norm between benign and adversarial examples.
In contrast to that, Papernot et al. \cite{7467366} proposed in their ``Jacobian-based Saliency Map Attack'' (\textit{JSMA}) to restrict the perturbations with respect to the $L_{0}$ norm.
Hence, the attack tries to minimize the amount of input points being changed rather than restricting the global change to the input.

This idea was picked up by Su et al. \cite{DBLP:journals/corr/abs-1710-08864}.
In this publication the authors successfully fooled DNNs using their \textit{One Pixel Attack}.

Currently the most powerful attack was proposed by Carlini and Wagner (\textit{C$\&$W}) in \cite{DBLP:journals/corr/CarliniW16a}.
This method is capable of crafting adversarial examples even for targets protected by state-of-the-art defense methods.
The basic idea of the attack is, instead of optimizing the loss function directly, to introduce a cost function $f_{y}$ as substitute.

Moosavi-Dezfooli et al. \cite{DBLP:journals/corr/Moosavi-Dezfooli16} presented \textit{Universal Adversarial Perturbations}. 
Rather than calculating individual adversarial examples, the authors calculated a universal perturbation such that when added to an arbitrary input, the target network is fooled.

If the attacker does not have access to the target model and its parameters, black-box attacks still pose an alternative to manipulate the classifications.
In this paper, we focus on \textit{Transfer Attacks} exclusively, when confronted with a black-box situation.
Here, the attacker uses a neural network over which she has full control and creates adversarial examples for it.
The attacker then transfers the resulting examples to the actual target to provoke a misclassification.

\subsection{Defenses against Adversarial Attacks}
Akhtar and Mian \cite{DBLP:journals/corr/abs-1801-00553} categorize adversarial defenses using three classes. 
The first class introduces a modified training procedure or various prepossessing methods to the input data, respectively.
In the second and third class, modifications to the targeted model itself or an additional model are introduced to increase overall robustness.
For both classes, some techniques aim to increase the robustness by detecting adversarial examples. 
As we propose a new technique to achieve the same goal, we sum up related methods into a fourth class. 
In the following we make use of this categorization and present previous findings for each class and explain them briefly.
We pay special attention to the methods trying to detect adversarial examples.
For further analysis of state-of-the-art detection techniques we refer to the survey by Carlini and Wagner \cite{Carlini:2017:AEE:3128572.3140444}.

\subsubsection{Changes to the Training Process or Input Data} \label{compression}
\textbf{Adversarial training:} The most intuitive and widely performed defense technique is to include adversarial examples in the training phase of the model to protect.
This is achieved by simply extending the training set with adversarial examples \cite{45816}.
Adversarial training is often introduced by authors of attacks as the first strategy to prevent a successful attack \cite{42503,43405,DBLP:journals/corr/Moosavi-Dezfooli15}. 

The authors of \cite{43405} proposed training based on a modified objective function.
The idea is to force the prediction of adversarial and benign images of one class to the same direction.
Additional regularization avoids over-fitting, which again increases the robustness of the network against unseen adversarial examples \cite{43405,DBLP:journals/corr/Sankaranarayanan17b}.

In 2017 Madry et al. \cite{DBLP:conf/iclr/MadryMSTV18} interpreted adversarial defense as a robust optimization problem. 
The authors claimed the PGD attack method to be a universal attack as it supposedly makes use of the local first order information about the target network in a superior way compared to other attack techniques.
Hence, the authors used examples created with PGD during the adversarial training. 
The resulting networks are robust against other adversaries, which is also shown in \cite{DBLP:journals/corr/abs-1709-10207}.

As adversarial training is easy to implement it may act as a first line of defense against known attacks. 
Nevertheless, it should not be used as the single approach to protect against adversaries.
Moosavi-Dezfooli et al. \cite{DBLP:journals/corr/Moosavi-Dezfooli16} showed that adversarially trained models are still vulnerable using other known attack methods.
Moreover, Tramèr et al. \cite{46638} presented a two-step attack method which also circumvents security provided by adversarial training.
The final drawback of adversarial training is the fact, that it is prone to black-box attacks \cite{DBLP:journals/corr/NarodytskaK16, Papernot:2017:PBA:3052973.3053009}.

\textbf{Data compression and feature squeezing:} 
Dziugaite et al. \cite{DBLP:journals/corr/DziugaiteGR16} first showed that adversarial images created with the FGSM method can be classified correctly if JPG compression is applied.
Based on this finding, further experiments using JPG and JPEG compression resulted in successful defense methods \cite{DBLP:journals/corr/abs-1711-00117, DBLP:journals/corr/DasSCHCKC17}.
However, Shin and Sing \cite{JPEG-resistant} showed that a considerable amount of adversarial images are not vulnerable to a JPEG compression, especially when crafted with the C$\&$W method.

Similar strategies have been proposed in \cite{DBLP:journals/corr/XuEQ17a} and \cite{DBLP:journals/corr/LiangLSLSW17}. 
Here ``Feature Squeezing'' is used to reduce the complexity of the inputs, by reducing the color depth or applying smoothing filters.
Additionally, adversarial example detection can be conducted which we will discuss in Section \ref{detection}.

The disadvantage of using the above mentioned techniques prior to the classification is a decreasing classification accuracy.
Since no prior knowledge about the images is given, each has to be compressed before being classified, resulting in a information loss for benign images. 

\textbf{Data randomization reprocessing:}
Luo et al. \cite{DBLP:journals/corr/LuoBRPZ15} proposed to apply the targeted neural network only to a certain region of the currently classified image.
This technique is shown to be a valuable countermeasures against adversarial images created by L-BFGS and FGSM based algorithms.
Xie et al. \cite{DBLP:journals/corr/XieWZZXY17} analyzed the effects of random resizing and padding.
The authors reported positive effects on the classification accuracy of adversarial images.
Similarly, Wang et al. \cite{DBLP:journals/corr/WangGZOXGL16} made use of a separately executed data-transformation module, which partially removes adversarial perturbations.

\subsubsection{Modifying the Network}
In recent years, a substantial increase of DL robustness was achieved by making changes directly to the network.
The inputs remain unchanged which reduces preprocessing time.

Gradient Hiding limits the accessibility of the gradients and successfully circumvents associated attacks.
Nonetheless, this technique does not provide protection against black-box attacks, as shown in \cite{Papernot:2017:PBA:3052973.3053009}. 

Related to Gradient Hiding, Ross and Doshi-Velez \cite{DBLP:journals/corr/abs-1711-09404} introduced Gradient Regularization.
The authors proposed to penalize the degree of variation of the output, based on changes in the input.
This concept led to further techniques like \cite{journals/corr/LyuHL15} and \cite{journals/corr/ShahamYN15}.

In 2015, Papernot et al. \cite{7546524} presented Defensive Distillation.
The originally introduced distillation technique shown by Hinton et al. \cite{44873} aims to simulate a neural network using a smaller one.
In contrast to that, the authors try to generate a smoother, less sensitive version of the original model.
This is achieved by reusing the probability vectors of the training data during the training of the model.
In 2017, Papernot and McDaniel further improved the concepts conveyed in the initial publication.
Nonetheless, Carlini and Wagner \cite{DBLP:journals/corr/CarliniW16a} claim their C$\&$W attack to be successful against Defensive Distillation.

\subsubsection{External Network Add-Ons}
Akhtar et al. \cite{DBLP:journals/corr/abs-1711-05929} proposed the idea of Perturbation Rectifying Networks (PRN). 
These sub-networks are added in front of the original network and are trained separately after the actual training phase.
The PRN rectifies the perturbations on the adversarial images, which are subsequently identified by an additional detector.

Since the 2014 released paper by Goodfellow et al. \cite{NIPS2014_5423}, Generative Adversarial Networks (GANs) are widely used and referred to in numerous publications.
Some promising publications using GANs to protect DNNs against adversarial attacks are \cite{DBLP:journals/corr/LeeHL17, DBLP:journals/corr/ShenJGZ17, DBLP:journals/corr/abs-1805-06605}.

\subsubsection{Detecting Adversarial Examples} \label{detection}
Our concept can be added to this class of defense strategies, hence, we provide a detailed overview of the latest related findings.
As mentioned  before, detection techniques can be based on both, changes to the input data or to the model itself. 
Additionally, observations of the model behaviour or the model's input provide insights on whether processed inputs are of adversarial nature or not.
In Section \ref{compression} we showed various preprocessing and compression methods which can be applied in order to reduce the effects of adversarial perturbations.
Moreover, these methods can additionally be used to detect attacks. 

Baluja and Fischer \cite{DBLP:journals/corr/BalujaF17} showed this by using the so-called feature squeezing technique: 
the authors created different versions of the input, based on different squeezing methods and let the target network classify them.
If the returned labels differ, the authors assume this input to be adversarial.
In a follow-up work, Xu et al. \cite{DBLP:journals/corr/XuEQ17a} used this technique to protect networks against the C$\&$W attack.

Similarly, Hendrycks and Gimpel \cite{DBLP:journals/corr/HendrycksG16b} performed a principal component analysis (PCA) on the inputs of neural networks.
The authors found that for adversarial examples, a higher weight is placed on larger principal components in comparison to benign examples.
With this knowledge, a binary classifier can detect attacks.

Liang et al. \cite{DBLP:journals/corr/LiangLSLSW17} interpreted adversarial perturbations as noise and tried to detect them by using scalar quantization and smoothing filters.

A more straightforward approach was evaluated by Gong et al. \cite{DBLP:journals/corr/GongWK17}. 
By applying a binary classifier on the input examples directly, the authors were able to detect adversarial input among benign examples.
Positive results were achieved using the MNIST dataset exclusively, during a later analysis in \cite{Carlini:2017:AEE:3128572.3140444} the approach failed to reach similar detection rates for different datasets.

Meng et al. \cite{Meng:2017:MTD:3133956.3134057} proposed their framework MagNet, which evaluates the original dataset and analyses the manifold of the benign examples.
If a new examples is passed to the network to be classified, it is compared to the findings about the manifold.
This method is shown to be vulnerable against attacks incorporating larger perturbations shown in \cite{DBLP:journals/corr/abs-1711-08478}.

A comparable pre-classification was introduced by Grosse et al. \cite{DBLP:journals/corr/GrosseMP0M17}.
The authors used the maximum mean discrepancy test, based on sets of benign and adversarial examples.
This test provides evidence on whether the two sub-datasets are drawn from the same distribution or not.

Hosseini et al. \cite{DBLP:journals/corr/HosseiniCKZP17} added a new class to the used dataset and try to unify adversarial examples in it. 
During training, the network is set to assign adversarial images to this so-called NULL class.

Metzen et al. \cite{metzen2017detecting} showed a method by adding a sub-network to the original neural network.
This sub-network is adversarially trained and acts as a binary classifier during the classification of the inputs.
In \cite{DBLP:journals/corr/LuIF17}, the authors showed that this method can again be bypassed by an attack.

Lu et al. \cite{DBLP:journals/corr/LuIF17} hypothesized that adversarial examples produce a pattern of Relu activation values in the late stages of a target network which differ from those based on benign examples.
In their framework called SafetyNet, the authors used a radial basis function support vector machine (SVM) to distinguish between original and perturbed examples.

Trying to increase the security of convolutional neural networks (CNNs), Li et al. \cite{DBLP:journals/corr/LiL16e} extracted the intermediate values after convolutional layers.
The authors performed a PCA of the extracted features and a cascaded classifier to detect attacks. 

In 2017, Feinman et al. \cite{Feinman2017DetectingAS} tried to detect adversarial attacks using two features which they extracted from dropout neural networks.
With these features a simple logistic regression is performed as the basis for a binary classifier. 
The first feature the authors introduced is the density estimate, based on which the distance between a given example and the sub-manifold of a class is quantified. 
For this purpose the authors used the feature space of the last hidden layer of the target network.
With their second feature, the Bayesian uncertainty estimate, the authors introduced an alternative feature to detect adversarial examples missed by the first feature.
Here, points shall be detected which lie in low-confidence regions of the original input space, indicating an attack.

Similar to our method Ma et al. \cite{Ma_NDSS19} detect attacks by observing the NN's hidden activations.
The authors identify two exploitation channels which form the basis of their detection approach.
By extracting provenance and value invariants attacks are detected using a one-class SVM.
Compared to this method, we propose a fully automated end-to-end system without further feature engineering steps

\section{Methodology} \label{Methodology}
In this section we introduce our main concept of detecting adversarial examples during classification time.
The core idea originates from our hypothesis as initially indicated in Section \ref{sec:introduction}:
Adversarial examples provoke the dense layer neuron coverage of neural networks to behave in such a distinctive manner that attacks become detectable by observing their activity patterns.
We provide a detailed description on how to expand and build upon this idea in the following.

\begin{figure*}[]
\centerline{\includegraphics[width=1.05\textwidth]{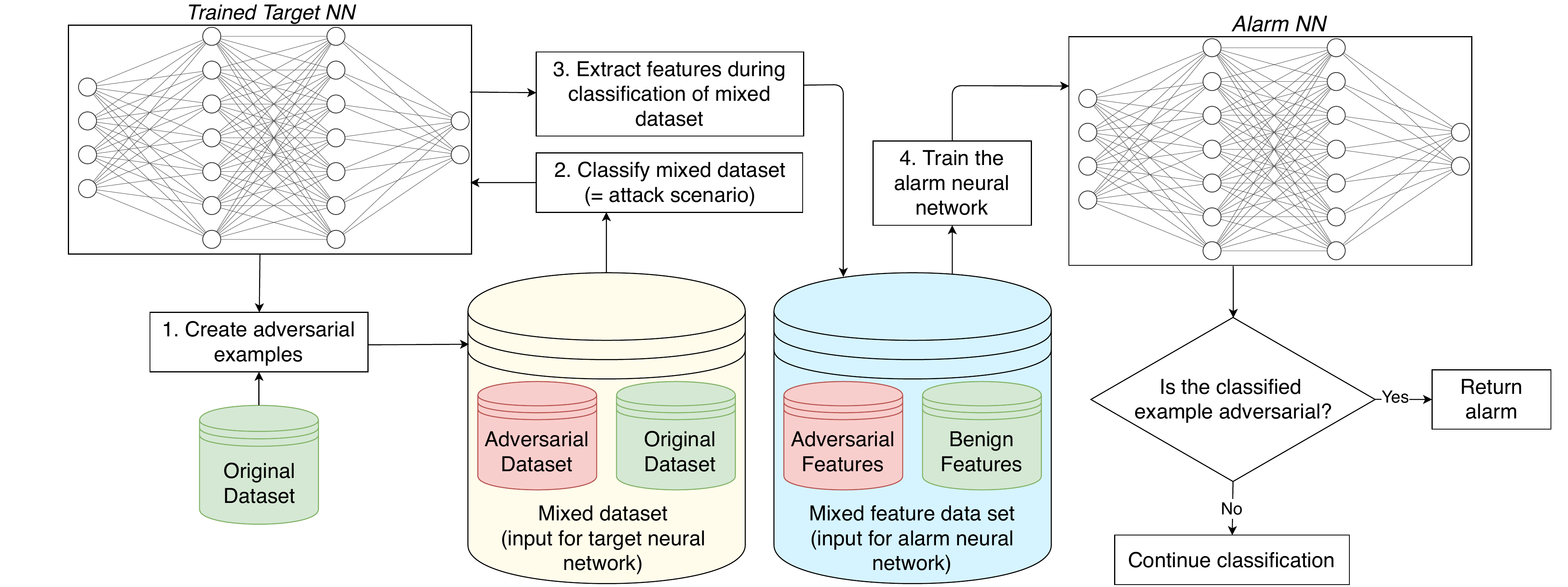}}
\caption{Overview of our concept showing the required neural networks, datasets, and calculations.}
\label{our_method}
\end{figure*}

Fig.~\ref{our_method} shows an overview of the overall process.
In the following, we discuss each step in detail. 
Joining these steps provides an end-to-end pipeline for fully automated detection of adversarial examples.

Our method is designed to help developers and maintainers of neural networks to secure their models against attacks.
Hence, we assume access to the fully trained model as well as (read-only) access to the benign training dataset $D_{benign}$.
We refer to the model that we want to secure as the \textit{target model} $N_{target}$. 
Our overall aim is to generate a secure model $N_{target}^{secure}$ that throws an alarm signal whenever an adversarial example is being processed. 
To achieve this, we first generate adversarial examples, second extract the dense layer neuron coverage, and third train an alarm model that enables secure operation of our initial target model.

\subsection{Generating Adversarial Examples}
In the first step of our concept, we generate adversarial examples $D_{adv}$ for our target model $N_{target}$. 
We craft these examples in a white-box manner by exploiting all available information.

It is important to note that we create adversarial examples for each class of the dataset.
Hence, we try to push the generated adversarial examples to be misclassified with an equal distribution among all remaining (i.e. false) classes.
This is a crucial step during the generation phase in order to cover all possible cases which might occur during the application of our method in the field.

Once the adversarial examples are generated, we summarize them in a dataset $D_{adv}$.
The outputs of the adversarial example generator, i.e., the elements of $D_{adv}$, are labeled as \textit{adversarial}, while the original unmutated samples $D_{benign}$ are labeled as \textit{benign}.

For the adversarial example generation, we use a wide range of adversarial crafting methods, including state-of-the-art techniques.
As we discussed in Section \ref{related_work}, the attacks do not only differ in success rates but also in their detectability.
Hence, by covering the currently strongest attacks we try to circumvent this issue.
Moreover, to cover the case of black-box attacks, we recommend to use transferred adversarial examples as well.
It is important to note that only these examples should be considered that actually provoke a misclassification of the target network.

\subsection{Extracting Dense Layer Neuron Coverage}
This step of our concept can be viewed as an additional preprocessing phase prior to the application of the target model in its intended environment.
We refer to this step as \textit{feature extraction}.
Here, the datasets \(D_{benign}\) and \(D_{adv}\) are fed to the well-trained target model which performs classifications using the individual samples.
Since the feature extraction is not part of the actual function and objective of $N_{target}$, we ignore its classification outputs.
Instead, we extract the activation values of all available dense layers and concatenate them to one sequence.
The resulting datasets, which hold the sequences for all samples, are called \(I_{benign}\) and \(I_{adv}\), respectively.
For further usage of the datasets, we adopt the labels to distinguish between adversarial and benign samples.
In summary, the dataset \(I_{<attackname>}\) holds the activation value sequences for all benign and adversarial examples regarding the target model for one specific attack method.
We preserve this separation of the activation value sequences, because we assume the different attack methods to have characteristic impacts on the behavior of the target and the resulting features.
This not only enables us to detect the individual attacks, but also to assess the impact of the individual crafting methods.

\subsection{Training an Alarm Model}
The dense-layer neuron coverage we extracted in the previous step builds the basis for our core idea to detect adversarial examples.
Assuming that this coverage contains information about the model, its behaviour, and the input sent to it, we need a supplementary analysis of the extracted information.

Previous work \cite{Feinman2017DetectingAS}, as discussed in Section \ref{detection}, followed a related idea.
The authors try to extract information from neural layers and further manually process them to detect adversarial images.
However, taking information directly from all dense layers of the trained model is more efficient for providing an end-to-end solution without further processing steps.

Accordingly, we propose the following method  which generalizes well over different scenarios and model architectures:
We interpret the analysis of the dense layer features as a binary classification.
Instead of including hands-on measures and distinguishing between different scenarios, we train an additional neural network to perform the required actions.
We refer to this network as the \textit{alarm model} $N_{alarm}$ in the following.

To train the alarm model, we use the features stored in \(I_{<attackname>}\).
Therefore, the network is trained to distinguish between activation values gathered during the classification of benign and adversarial features, respectively.
In the actual secure operation phase, $N_{alarm}$ performs a binary classification of newly extracted features provoked by the current samples to classify.
This forms the final adversarial example detection running alongside $N_{target}$.

The architecture of the alarm model heavily influences the success of our approach.
Furthermore, different architectures need to be tested against each other to provide a viable solution.
In Section \ref{Experiments}, we give a recommendation for a specific architecture.
Still, future work needs to further evaluate this part of the concept.

Note that we recommend to create one alarm model for each introduced attack method.
The attack methods differ in their approach and complexity and thus influence the neuron activation patterns distinctively.
Hence, using a set of different alarm models allows us to detect a broader range of attacks.
Furthermore, we are able to evaluate the capability of each alarm model version of detecting different attack methods.
This provides information on the applicability of our concept when detecting future attack methods.

\subsection{Concept Overview}
After we introduced the building blocks of our concept, we can now link them and present an overview of our approach with Algorithm \ref{main_algo}.
The application of our method in a real-world scenario can be divided into two steps. 
A prior initialization phase prepares our framework to enable a secure operation of the target model $N_{target}^{secure}$.

In the initialisation phase, we create adversarial examples and perform the according feature extraction steps.
We have shown the importance of using different attack methods to create the adversarial examples.
This ultimately leads to an assemblage of alarm models, each capable of detecting adversarial examples created by one specific attack method.

During the secure operation of the target model, we continuously extract the features during classification of new, unseen samples.
The resulting activation sequences are fed to all available alarm models to perform binary classifications.
If the classifications indicate an attack, our framework throws an alarm signal and a human expert is consulted to evaluate the current input.
Here, the maintainer chooses if one assumes an attack based on one or more alarm signals, majority votes, or all alarm models synchronously indicating such an event.

\begin{algorithm}[]
    \KwIn{$D_{benign}$, $D_{test}$, $N_{target}$, $N_{alarm}$}
    \KwResult{$N_{target}^{secure}$}
    \For{Initialization}{
        $D_{adv}$ $\leftarrow$ CreateAdvExamples($D_{benign}$, $N_{target}$) \;
        $I_{benign}$ $\leftarrow$ ExtractInformation($D_{adv}$, $N_{target}$) \;
        $I_{adv}$ $\leftarrow$ ExtractInformation($D_{benign}$, $N_{target}$) \;
        $N_{alarm}$ $\leftarrow$ Train($N_{Alarm}$, $I_{benign}$, $I_{adv}$) \;
    }
    \While{Secure Operation}{
        \While{1}{
            $x$ $\leftarrow$ Sample($D_{test}$) \;
            $y_{target}$ $\leftarrow$ Classify($N_{target}$, $x$) \;
            $i_{y_{target}}$ $\leftarrow$ ExtractInformation($x$, $N_{target}$) \;
            $y_{alarm}$ $\leftarrow$ Classify($N_{alarm}$, $i_{y_{target}}$) \;
                \If{$y_{alarm}$ == 1}{
                    Alarm \;
                    ConsultHumanExpert();}
        }
    }
    \caption{Main algorithm, divided in an initialization and a secure operation phase.}
    \label{main_algo}
\end{algorithm}

\section{Implementation and Experimental Setup} \label{Experiments}
In the following, we present details regarding our proof-of-concept implementation and our experimental setup.
We discuss our choice of datasets and accompanying target model architectures on which we tested our approach.
The description of feature extraction and the following alarm model training form the core of this section.
We introduce one exemplary alarm model implementation to finally detect adversarial examples.
Subsequently we present the test scenario for which we show the evaluation results in the next section.
This includes details regarding our test environment.
Finally, we introduce a series of extension test scenarios which are not part of the actual proof-of-concept.
Instead, we try to provide the reader with a better intuition for our approach and rule out possible restrictions.
We motivate generality of our main hypothesis as discussed in Section \ref{sec:introduction} and the resulting idea of this paper.
For this purpose, we included tests for noisy images, two special target model architectures, and two additional dataset types, namely a natural language and an audio dataset.
Moreover we evaluate adaptive attacks, in which the attacker has full knowledge and control over the target and alarm model.

In this section of the paper, we solely covey information on the executed experiments.
Hence, we refer to the next chapter for the respective results.
We divided the proof-of-concept into two sections in order to preserve the paper's readability.

\subsection{Datasets}
For proof-of-concept, we considered the MNIST \cite{deng2009imagenet} and CIFAR10 \cite{krizhevsky2009learning} image datasets.
We decided to do so based on the following two observations:
First, to allow a comparison of our method and state-of-the art defense techniques.
Second, the usage of image datasets enables us to better visualize the adversarial examples and evaluate the performance of different attack methods.

The MNIST dataset consists of \num{70000} handwritten digits ranging from \num{0} to \num{9} of which \num{60000} build the training set and \num{10000} the test set.
Each digit is represented by $\num{28}\times\num{28}$ grey-scale pixels.

CIFAR10 consists of \num{60000} colored images of which again \num{10000} images build the test set.
Each image is stored using $\num{32}\times\num{32}\times\num{3}$ pixels, which makes this dataset more difficult to classify.
The CIFAR10 dataset therefore enables us to perform tests close to a real-world scenario.

To prove detectability of adversarial examples in the natural language processing (NLP) context, we used the IMDB dataset of movie reviews \cite{maas-EtAl:2011:ACL-HLT2011}. 
Both, the train and test set, contain \num{25000} samples each.
For both subsets, positive and negative reviews are distributed evenly.

The audio examples we considered during our tests are drawn from the Mozilla Common Voice dataset \cite{mozilla_common_voice}, which contains \num{803} hours of recorded human sentences.
Contrary to the above mentioned datasets, the instances in the Common Voice dataset are used for speech-to-text conversions rather than being classified according to known classes.

\subsection{Target Models}
Throughout the proof-of-concept, we payed attention to use state-of-the art target models in order to keep a close relation to real-word scenarios.
Table \ref{target_models} sums up the used models and shows their training and test accuracy as well as a short summary on the individual architectures.
For MNIST, we chose LeNet \cite{LeCun:1989:BAH:1351079.1351090} and a simple Multi-Layer-Perceptron (MLP) \cite{kerasEx_MNIST} that we refer to as kerasExM.   
In contrast, for CIFAR10 we decided to utilize ResNet \cite{ResNet_CIFAR10} and a deep CNN \cite{kerasEx_CIFAR10}, we refer to as kerasExC.

Furthermore, in order to evaluate if our method can generally be applied to a wide range of DL architectures, we conducted experiments using the two following examples.
On the one hand, we included a Long Short Term Memory (LSTM) based target model.
This architecture achieves a remarkable performance, especially in image classification tasks.
On the other hand, we included a so-called Capsule Network (Capsule NN), which is supposed to be more robust than conventional models.

For our tests regarding the NLP dataset, we again used an LSTM target model. 
We chose a pre-trained DeepSpeech implementation (0.4.1) \cite{mozilla_deep_speech} during our tests on audio examples.
Hence, we did not add its training and test accuracy to Table \ref{target_models}.
This neural network converts speech in the form of audio files to according text.

\begin{table*}[h]
\centering
\caption{Target models used during experiments. Showing details on the architectures and performance. Capsule and ResNet are trained using the \textit{adam} optimizer \cite{kingma2014adam}. The remaining models are trained with \textit{stochastic gradient descent}. The DeepSpeech model is pre-trained.}
\begin{tabular}{|l|l|>{}m{4cm}|l|l|}
\hline
\textbf{Dataset}         & \textbf{Model Name} & \textbf{Model Details} & \textbf{Training Accuracy} & \textbf{Test Accuracy} \\ \hline
\multirow{4}{*}{MNIST}   & LeNet \cite{LeCun:1989:BAH:1351079.1351090}              & -- 2 convolutional layers with filter size 5 \newline -- each convolutional layer is followed by a max-pooling layer with size 2 \newline -- 2 dense layers after each max-pooling layer    & 0.976                      & 0.987                \\ \cline{2-5}
                         & kerasExM \cite{kerasEx_MNIST} & -- one hidden layer with 512 neurons 
& 0.972                      & 0.985                \\ \cline{2-5} 
                         & CapsuleNN \cite{sabour2017dynamic}             & -- 10 capsules each of size 6
               & 0.992                      & 0.991                  \\ \cline{2-5}
                         & LSTM\cite{hochreiter1997long}                &     -- 1 LSTM layer followed by two dense layers with 64 and 32 neurons
 & 0.975                      & 0.978                  \\ \hline
\multirow{2}{*}{CIFAR10} & kerasExC \cite{kerasEx_CIFAR10} &    
 -- 4 convolutional layers with filter of size 3 \newline -- each pair of convolutional layers followed by a max-pooling layer of size 2 \newline -- last hidden layer of dimension 512 is fully connected.
 & 0.888                      & 0.889                  \\ \cline{2-5}
                         & ResNet \cite{ResNet_CIFAR10}             & -- 3 blocks followed by an average pooling size of 8
                         & 1.000                      & 0.840                  \\ \hline
IMDB                     & LSTM (for NLP)                                         & -- 1 embedding layer \newline -- 1 \num{64}-neuron dense layer
& 0.996                          & 0.81                    \\ \hline
Mozilla Common Voice     & DeepSpeech \cite{mozilla_deep_speech}                                  &  -- containing 2 parts \newline -- convolutional and recurrent network
 & --                          & --                    \\ \hline 
\end{tabular}
\label{target_models}
\end{table*}

\subsection{Adversarial Attack Methods}
\begin{figure}[h]
\centerline{\includegraphics[width=0.6\textwidth]{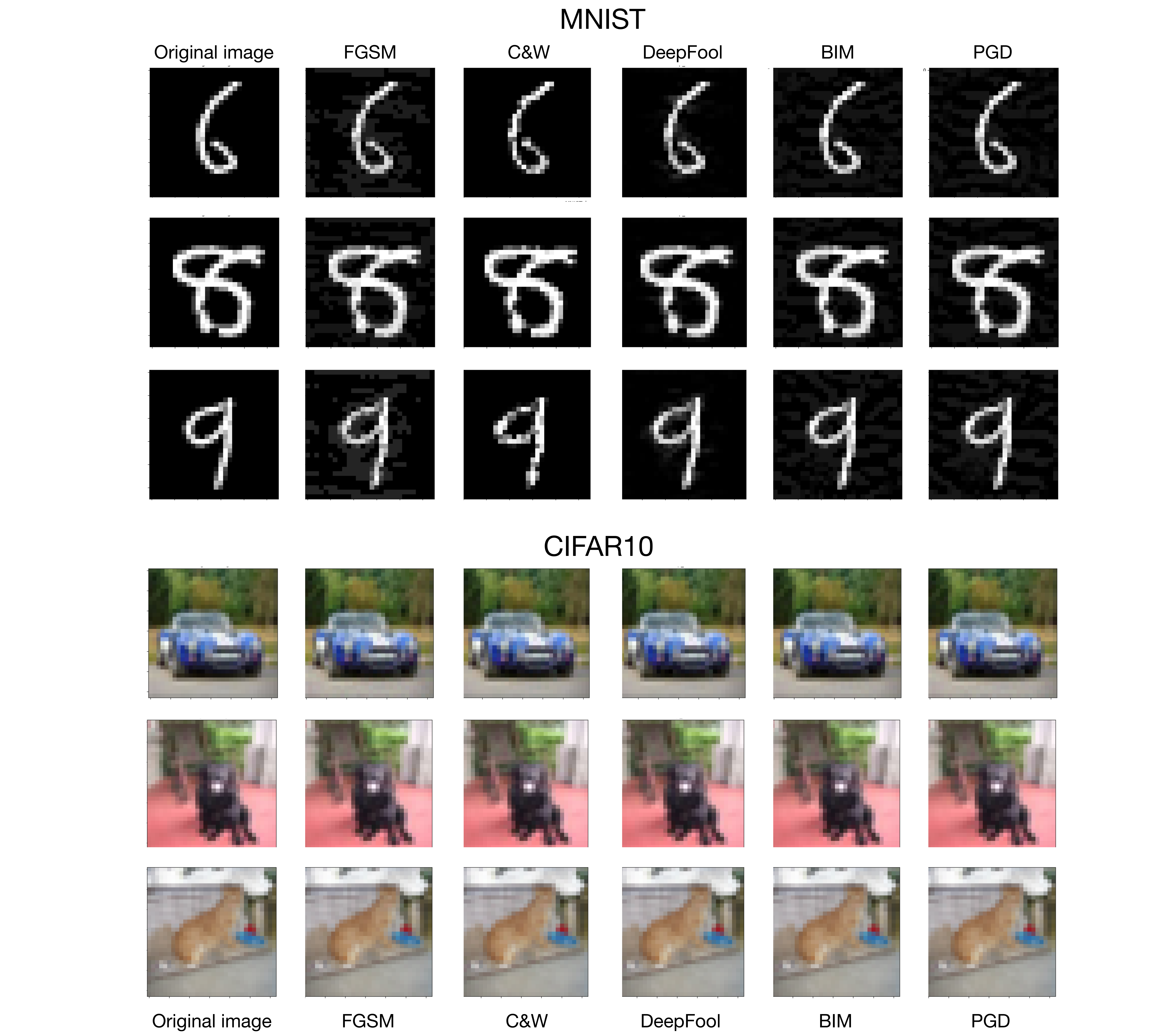}}
\caption{Adversarial images created  with (from left to right): FGSM, C$\&$W, DeepFool, PGD, BIM. The top images are based on the MNIST dataset and are crafted on the target model LeNet. The bottom images are based on the CIFAR10 dataset and are crafted on the target model kerasExC.}
\label{adversarial_images}
\end{figure}

We evaluate the detectability of the following attack methods: FGSM, C$\&$W, DeepFool, PGD, and BIM.
The motivation to do so originates from the nature and popularity of these methods.
We include diverse attacks, such that remarkable differences in the basic idea can be seen.
Moreover, we payed attention to add attacks which differ in strength and complexity.
The C$\&$W attack, for instance, is currently considered to be the most powerful attack.
Hence, this and future adversarial detection schemes need to be tested against this method.
Fig.~\ref{adversarial_images} shows a series of adversarial images for both datasets crafted with the above mentioned techniques.

Alongside the five stated attack methods, we include one \textit{black-box} attack as well.
We create adversarial images in a \textit{white-box} setup on model \textit{A} and try to fool model \textit{B} with the resulting examples. 
We call this \textit{transfer} attack in the following.

Since the actual crafting and implementation of the attacks is not part of our concept, we applied the \textit{foolbox} framework \cite{rauber2017foolbox} to generate adversarial examples for MNIST and CIFAR10.
To create adversarial examples based on the Common Voice dataset, we refer to \cite{8424625}.
Finally, for the IMDB dataset we created an algorithm to create adversarial examples, which we briefly describe in Appendix \ref{appendix_d}.
In future work, we will further explore and refine this attack method.

\subsection{Alarm Model Architecture and Training}
During our proof-of-concept, we exclusively used one specific alarm model architecture to detect adversarial examples.
We chose to do so in order to show the generality of our concept and keep the space of tunable parameters as small as possible.
Hence neither the used dataset nor the applied target model, which needs to be protected, affect our alarm model architecture. 
For future work or in real-world applications, a deliberate choice of the alarm model will likely further improve the strength of our concept.
In this paper, we use a DNN with six dense layers which we train for ten epochs and a batch-size of \num{100} in each scenario.
We use the \textit{adam} optimizer \cite{kingma2014adam} during training.
Note that in some cases this alarm model suffers from underfitting.
We included a more detailed description of the architecture in Appendix \ref{appendix_a}.

\subsection{Main Test Scenario} \label{test_scenario}
We have discussed all building blocks of the experimental setup and the according preliminaries, in this section, we focus on the actual test scenario and performed actions.
We assume ourselves in the position of the trained model's maintainer and try to increase its trustworthiness.
Each step we present is performed for all introduced attack methods.
For the sake of simplicity we neglect this fact in the following and show each step once.

In the first step, we craft adversarial images using the above stated methods.
Since we have full control over $N_{target}$ and its parameters, we perform \textit{white-box} attacks.
During this process, we pay attention to the way the datasets have been split beforehand.
Consequently, we create two separate adversarial datasets, based on the train and test subsets, respectively.
By doing so, we simulate the real-world case in which the maintainer only possesses the training data.
The samples in the test set simulate inputs fed to the target during its usage in the field.
Additionally, we can later check detectability of adversarial examples which are based on unseen benign inputs to rule out a detection bias.
We refer to the datasets as $D_{adv}^{train}$ and $D_{adv}^{test}$, respectively.
For both, MNIST and CIFAR10, we create adversarial counterparts of $D_{benign}^{train}$ and $D_{benign}^{test}$ containing a similar amount of examples:
(\num{60000}, \num{10000}) for MNIST and (\num{50000}, \num{10000}) for CIFAR10. 
We let $N_{target}$ classify all samples in the four datasets and store the activation sequences in $I_{benign}^{train}$, $I_{benign}^{test}$, $I_{adv}^{train}$, and $I_{adv}^{test}$.
Each individual set contains features extracted during the classification of benign and adversarial samples while we preserve the division between test and training samples.
Note that we neglected this split of the datasets in Algorithm \ref{main_algo} for the sake of simplicity.

In the second step, we use $I_{adv}^{train}$ and $I_{benign}^{train}$ to train the alarm model.
Hence, for each target model and attack method, we create one specific alarm model which we call $N_{alarm}$.
To test our capability of detecting adversarial examples, we let $N_{alarm}$ classify all samples in $I_{benign}^{test}$ and $I_{adv}^{test}$.

To further show the generality of our concept, we conduct cross-testing cases.
This means that we test a trained alarm model against the features of a different attack.
With this, we simulate the scenario in which we encounter a new and yet unknown attack.

Furthermore, we create a combined alarm model $N_{alarm}^{combined}$ which is trained using all features gained during the extraction of a specific target model.
Here, we verify if considering more information, based on a wider range of attacks, improves the alarm model's performance and provides a stronger setup in the context of our concept.

\subsection{Additional Experiments and \\ Adaptive Attacks} \label{further_experiments}
At the beginning of this section, we briefly introduced the supplementary tests we conducted to further establish confidence in our approach.
These tests can be divided into four parts.

In the first part, we used the previously created adversarial images for the MNIST and CIFAR10 datasets and conducted transfer attacks targeting an LSTM neural network and a Capsule network, respectively.
This experiment gives insights on whether our approach can be applied in the context of different DL architectures or not.

In the second part, we run two experiments on the regular target models classifying MNIST and CIFAR10 images.
With the first test we analyze if our concept is robust to noisy input images.
Hence, we exclude the possible effect in which our framework is solely able to distinguish between clean, benign images and adversarial images containing noise.
We achieve this by creating noisy benign images with the same amount of distortion as their adversarial counterparts.
For this purpose we calculate the distances between the original and adversarial images with respect to the used distance metric of the attack.
The resulting datasets now contain original, adversarial, and benign noisy images.

Subsequently, we provide evidence for our initial hypothesis of the paper presented in Section \ref{Methodology}.
We analyze the dense layer activation values of the target models during misclassification of original inputs.
Thus, we extract the according features and treat them as activation values gained during classification of adversarial examples.
The features are then used to train an alarm model which tries to detect misclassifications during testing of the benign dataset.

In the third part, we test our concept in the context of two additional types of datasets.
We investigate if we are able to detect adversarial examples in NLP and audio datasets.
This test gives first evidence on the applicability in numerous DL-based environments.
State-of-the-art defense methods mostly focus on image processing target models.
Therefore, proving the applicability of our concept in additional types of datasets poses a significant step towards more robust defense methods.
We introduce the test environment and results in a stand-alone paragraph in Section \ref{nlp_audio_tests}.

Finally we evaluate adaptive, white-box attacks in which the attacker has perfect knowledge of the target model and our detection method.
As shown by Carlini and Wagner \cite{Carlini:2017:AEE:3128572.3140444}, the majority of proposed detection methods can easily be bypassed by an adaptive attack.
To perform such an attack, we require a loss function based on the new target $N_{secured}$, which consists of the original target model $N_{target}$ and our alarm model $N_{alarm}$.
We use the loss function defined by Carlini and Wagner \cite{Carlini:2017:AEE:3128572.3140444}, which the authors use to attack similar detection-based defense methods.
Here, we profit from the code the authors published to reproduce their results \cite{CarliniAdaptiveAttacks}.
Thus, we generate adversarial examples for $N_{secured}$, using the C\&W method exploiting perfect knowledge of the target and alarm model.
We perform this attack on the secured version of LeNet for MNIST and on the secured version of kerasExC for CIFAR10.

\section{Evaluation} \label{Results}
In this section, we sum up the evaluation results.
We split this into three parts.
First, we discuss our analysis of the extracted features and show their distribution for one specific example.
Second, we present the main results accumulated during our experiments.
This includes the performance of the different alarm models while detecting adversarial examples.
Third, we present the results we gained during our supplementary experiments including adversarial example detection in NLP and audio datasets and adaptive attacks.

\subsection{Feature Analysis}
\begin{figure*}[]
\centerline{\includegraphics[width=.9\textwidth]{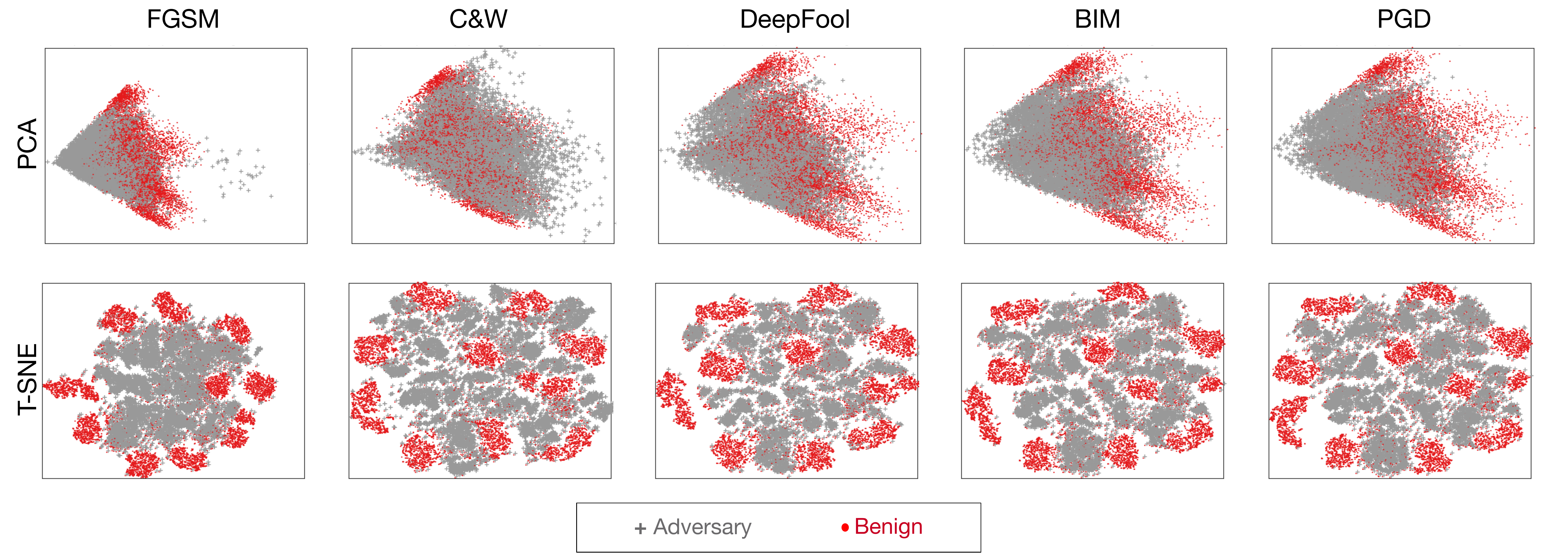}}
\caption{Visualization of the extracted features during the classification of MNIST-based adversarial and benign images for the LeNet target model. For better visualisation, the dimensionality of the features were reduced using PCA and t-SNE, respectively. Each column shows the plots for one attack method. The red dots and gray crosses represent the benign and adversarial samples, respectively.}
\label{extracted_features}
\end{figure*}

As the extracted features are the core of our hypothesis and concept, we illustrate some findings during the analysis.
In Fig.~\ref{extracted_features} we show neuron activation sequences for the LeNet target model for all attack methods after we reduced the dimensionality of the data using PCA and t-distributed stochastic neighbor embedding (t-SNE).
The figures show the neuron coverage of the dense layers during the classification of benign and adversarial images.
The red dots and gray crosses represent the benign and adversarial instances.

We can clearly see a difference in the dense-layer activation values.
Interestingly, we can see artifacts of the ten classes of the MNIST dataset in the t-SNE figures.
This finding gives first evidence on the verity of our initial hypothesis.
Furthermore, we can show a first estimate for the complexity and detectability of the attack methods.
The PCA data points of the C$\&$W-based activation sequences overlap to a higher extent than for the remaining methods.
This indicates a more challenging detection of the C$\&$W attack.

Nonetheless, since we want to provide an end-to-end framework to detect adversarial examples, we directly use the raw extracted data.

\subsection{Detection Performance} \label{main_results}
In this section, we present the performance of our concept using the differently trained alarm models. 
We assess the success of our detection method with the f1-score.
Furthermore we present the the mean \textit{false positive} and \textit{false negative} rates.

During the proof-of-concept, we conducted numerous experiments.
For the sake of simplicity and readability, we exclusively include results significant for the proof of our main idea and hypothesis.
We provide the remaining results, tables, confusion matrices and figures in the appendix of this paper.

In Table \ref{simple_results}, we list the f1-scores of the individual alarm models when tested against their dedicated attack method.
We can see a strong detection capability for all attack methods and target models regarding the MNIST dataset.
The respective f1-scores range above \num{0.9}.  
For the CIFAR10 dataset our framework detects the majority of examples and poses a viable solution for real-world applications.

\begin{table*}[h]
\centering
\caption{F1-Scores of the individual alarm models when tested against their underlying attack method. Main results of the evaluation.}
\begin{tabular}{|l|l|p{1.5cm}p{1.5cm}p{1.5cm}p{1.5cm}p{1.5cm}|}
\hline
\multirow{2}{*}{\textbf{Dataset}} & \multirow{2}{*}{\textbf{Target Model}} & \multicolumn{5}{l|}{\textbf{f1- Score of the Alarm Models with the according attacks:}}                                  \\ \cline{3-7} 
                                  &                                        & \textit{\textbf{FGSM}} & \textit{\textbf{C\&W}} & \textit{\textbf{DeepFool}} & \textit{\textbf{PGD}} & \textit{\textbf{BIM}} \\ \hline
\multirow{2}{*}{MNIST}            & LeNet                                  & 0.988                  & 0.977                & 0.981                       & 0.991                 & 0.990                 \\
                                  & kerasExM                                & 0.992                  & 0.975                & 0.982                       & 0.993                 & 0.991                 \\ \cline{1-7} 
\multirow{2}{*}{CIFAR10}          & kerasExC                                & 0.847                  & 0.733                & 0.855                       & 0.843                 & 0.852                 \\
                                  & ResNet                                 &   0.815              & 0.727                & 0.833                       & 0.833                 & 0.832                 \\ \hline
\end{tabular}
\label{simple_results}
\end{table*}

\begin{table*}[h]
\centering
\caption{F1-Scores of the combined alarm models when tested against each attack separately. The right most column holds the f1-scores of the combined alarm models when tested with a combined dataset containing all used attack methods.}
\begin{tabular}{|l|l|lllllll|}
\hline
\multirow{2}{*}{\textbf{Dataset}} & \multirow{2}{*}{\textbf{Target Model}} & \multicolumn{7}{l|}{\textbf{f1- Score of the combined Alarm Model with:}}                                                                                                          \\ \cline{3-9} 
                                  &                                        & \textit{\textbf{FGSM}} & \textit{\textbf{C\&W}} & \textit{\textbf{Deep Fool}} & \textit{\textbf{PGD}} & \textit{\textbf{BIM}} & \textit{\textbf{Transfer}} & \textit{\textbf{Combined}} \\ \hline
\multirow{2}{*}{MNIST}            & LeNet                                  & 0.981                  & 0.973                & 0.974                       & 0.981                 & 0.981                 & 0.931                      & 0.993                      \\
                                  & kerasExM                                & 0.984                  & 0.972                & 0.977                       & 0.984                 & 0.984                 & 0.947                      & 0.993                      \\ \cline{1-9} 
\multirow{2}{*}{CIFAR10}          & kerasExC                                & 0.771                  & 0.741                & 0.771                       & 0.772                 & 0.772                 & 0.754                            & 0.932                      \\
                                  & ResNet                                 & 0.812                        & 0.687
                     & 0.818                            & 0.820              & 0.819                      & 0.791                           & 0.864                           \\ \hline
\end{tabular}
\label{combined_alarm}
\end{table*}

To evaluate if a combination of features while considering different attack methods at once improves the performance of our method, we created a combined alarm model.
Hence, for each target model, its combined alarm model is trained with features extracted during the evaluation of all attack methods.
Table \ref{combined_alarm} provides an overview of this experiment.
The f1-scores show that the combined alarm models are able to detect all tested adversarial attack methods.
Comparing the results to Table \ref{simple_results}, we can report an superior performance of this combined method.

Above, we indicated the evaluation of cross-tests.
This corresponds to analyzing if the concept is capable of detecting new, unseen adversarial attacks.
In summary, for both datasets and for each target model we tested seven alarm model versions against six attack methods.
Five of the seven alarm models are based on the attack methods FGSM, C$\&$W, DeepFool, PGD, and BIM.
Two additional alarm models are the combined one, and the alarm model trained using features extracted during transfer attacks.
The result-space of the cross-tests exceeds the frame of this paper.
Hence, with the following two tables, we express the performance of our approach during cross-tests with mean values.
We added the individual result values to Appendix \ref{appendix_b}.
Table \ref{cross_testing_1} shows the mean performance of the individual alarm models, when tested against all attack methods.
In Table \ref{cross_testing_2}, we show the mean detectability of the individual attack methods, when utilizing all alarm models (except the model based on transfer attacks).
With the results we show that we are able to detect adversarial examples, for which the underlying method has not been known beforehand.
Thus, our method most likely enables us to detect even future, yet unknown attacks.

\begin{table*}[h]
\centering
\caption{Mean f1-scores of the individual alarm models when detecting all attack methods. }
\begin{tabular}{|l|l|lllllll|}
\hline
\multirow{2}{*}{\textbf{Dataset}} & \multirow{2}{*}{\textbf{Target Model}} & \multicolumn{7}{l|}{\textbf{Mean f1- Score of the individual Alarm Models with all attacks, model name:}}                                                                                                                               \\ \cline{3-9} 
                                  &                                        & \textit{\textbf{FGSM}} & \textit{\textbf{C\&W}} & \textit{\textbf{DeepFool}} & \textit{\textbf{PGD}} & \textit{\textbf{BIM}} & \textit{\textbf{Transfer}} & \textit{\textbf{Combined}} \\ \hline
\multirow{2}{*}{MNIST}            & LeNet                                  & 0.918                        & 0.945                      & 0.940                            & 0.878                       & 0.885                       & 0.950                            & 0.970                            \\
                                  & kerasExM                                & 0.892                        & 0.962                      & 0.943                            & 0.875                       & 0.888                       & 0.959                            & 0.975                            \\ \cline{1-9} 
\multirow{2}{*}{CIFAR10}          & kerasExC                                & 0.737                            & 0.626                          & 0.728                                & 0.727                           & 0.728                           & 0.729                                & 0.763                                \\
                                  & ResNet                                 & 0.733                            & 0.723                          & 0.747                                & 0.735                           & 0.737                           & 0.584                                & 0.791                               \\ \hline
\end{tabular}
\label{cross_testing_1}
\end{table*}

\begin{table*}[h]
\centering
\caption{Mean detectability of the individual attack methods, when tested against all alarm models expressed with the accroding f1-scores.}
\begin{tabular}{|l|l|p{1.5cm}p{1.5cm}p{1.5cm}p{1.5cm}p{1.5cm}p{2.5cm}|}
\hline
\multirow{2}{*}{\textbf{Dataset}} & \multirow{2}{*}{\textbf{Target Model}} & \multicolumn{6}{l|}{\textbf{Mean f1-Score of the individual attacks when tested with all alarm models}}                                                    \\ \cline{3-8} 
                                  &                                        & \textit{\textbf{FGSM}} & \textit{\textbf{C\&W}} & \textit{\textbf{DeepFool}} & \textit{\textbf{PGD}} & \textit{\textbf{BIM}} & \textbf{Transfer Attack} \\ \hline
\multirow{2}{*}{MNIST}            & LeNet                                  & 0.978                  & 0.770                & 0.940                       & 0.982                 & 0.982                 & 0.827                    \\
                                  & kerasExM                                & 0.987                  & 0.768                & 0.917                       & 0.987                 & 0.980                 & 0.832                    \\ \cline{1-8} 
\multirow{2}{*}{CIFAR10}          & kerasExC                                & 0.813                  & 0.500                & 0.803                       & 0.807                 & 0.805                 & 0.528                        \\
                                  & ResNet                                 & 0.802                  & 0.504                & 0.806                       & 0.807                 & 0.806                 & 0.685                        \\ \hline
\end{tabular}
\label{cross_testing_2}
\end{table*}

In the following we present the mean error rates during our evaluation.
The individual values can be found in Appendix \ref{appendix_c}.
For MNIST, the mean \textit{false positive} and \textit{false negative} ratios are \num{0.02} and \num{0.01}, respectively.
Similarly, for CIFAR10 we report the mean ratios of \num{0.22} and \num{0.16}, respectively.
For both examples we observe a higher \textit{false positive} ratio.
Hence, our method does not miss a disproportionate amount of adversarial examples. 
This is an important finding with regard to the applicability in a real-world setup.

\subsection{Results of further Experiments}
During our tests with a Capsule and an LSTM target network, we were able to detect adversarial images based on the MNIST dataset with f1-scores of \num{1.000} and \num{0.968}, respectively.
The positive results emphasize the applicability of our concept on a wide range of neural networks using dense layers.

In Table \ref{noise_tests}, we show the results of the tests containing noisy images.
The performance of the individual alarm models are decreased by \num{10}$\%$ in the worst case.
Hence, our method is robust against noisy inputs.
\begin{table}[h]
\centering
\caption{Performance of our method when detecting adversarial examples among clean and noisy benign images. Our capability of detecting attacks is not reduced significantly.}
\begin{tabular}{|l|l|l|l|}
\hline
\textbf{Dataset}       & \textbf{Target Model} & \textbf{Attack} & \textbf{f1-score} \\ \hline
\multirow{2}{*}{MNIST} & \multirow{2}{*}{LeNet}     & FGSM                   & 0.896             \\
                       &                            & C$\&$W                   & 0.913             \\ \hline
CIFAR10                & ResNet                     & FGSM                   & 0.791             \\ \hline
\end{tabular}
\label{noise_tests}
\end{table}

As the core hypothesis of our paper is strongly correlated to adversarial inputs, we introduced the tests regarding misclassified images.
We argued that the adversarial inputs provoke distinctive neuron coverage patterns, enabling detecting.
This excludes ordinary misclassifications of benign images due to inaccuracies in the model.
Tests in which we treated a misclassification the same way we treated attacks, provide evidence for our hypothesis.
In the tests we were not able to detect regular misclassifications with our framework.

\subsection{Experiments on Text and Audio Datasets} \label{nlp_audio_tests}
With the following test, we evaluate the generalizability of our method to different application domains.
During this analysis, we craft adversarial examples based on an NLP and audio dataset respectively.
As previously introduced, we use the IMDB and Mozilla Common Voice datasets.
Subsequently, we assess whether our framework is able to detect adversarial examples when classified by the according target models.

The process of generating adversarial examples for the two datasets is not part of the contribution of this paper.
Nevertheless, some basic notes are worth mentioning.
Regarding the IMDB dataset we use Algorithm \ref{alg_for_nlp_adv} found in Appendix \ref{appendix_d} to generate misclassified movie reviews.
Instead of adding or deleting words, we chose to replace words in the individual instances.
With this approach, we preserve the lengths of the classified sentences and reduce the distance between benign and adversarial examples.
We are able to report a detection f1-score of \num{0.960} for this dataset.

For the audio dataset, we used Carlini and Wagner's \cite{8424625} approach to create adversarial examples.
Extracting the features of the audio files leads to neuron activation sequences of different lengths.
This is the result of the various sampling rates during the recording of the original samples in the dataset.
To be able to perform binary classifications using all feature instances, we used a different alarm model architecture here.
The alarm model contains one LSTM layer followed by one output layer with two neurons to enable the detection of attacks.
For this dataset we are able to detect adversarial examples with an f1-score of \num{0.820}.

We clearly see successful detection of the majority of adversarial examples for the here used NLP and audio processing target models.

\subsection{Evaluation of Adaptive Attacks}\label{adaptive_attacks_evaluation}
To evaluate the robustness of our method against adaptive attacks, we consider the mean $L_{2}$-distance between adversarial and benign images to either fool $N_{target}$ or $N_{secured}$, respectively.
Hence, we generate adversarial images for $N_{target}$ and $N_{secured}$ using the same attack parameters to preserve comparability.
We list the chosen parameters in Appendix \ref{appendix_e}.
For MNIST and the LeNet target model, the mean $L_{2}$-distance is \num{61.82}$\%$ higher when attacking $N_{secured}$, compared to attacking the unsecured target model.
Regarding related work, this poses a significant improvement.
Carlini and Wagner \cite{Carlini:2017:AEE:3128572.3140444} show that related defense techniques only reach a \num{10}$\%$ higher mean $L_{2}$-distance.
The authors also show that related defense techniques can be bypassed with a success rate of \num{100}$\%$.
During our evaluation we reduce this rate to \num{99}$\%$.
For CIFAR10 and the kerasExC target model we achieve even better results. 
When attacking $N_{target}$ the mean $L_{2}$-distance is \num{0.14}.
In contrast to that, when attacking $N_{secured}$, the adversarial images show a mean $L_{2}$-distance of \num{5.06} with respect to their benign counterparts.
Moreover, only \num{84.1}$\%$ of adversarial examples of the adaptive attack are successful. 
Hence, we can report a significant security improvement of our target DNNs, even during white-box adaptive attacks.

\section{Discussion} \label{Discussion}
With the in-depth experiments in this paper we show the importance of including an analysis of the dense layer neuron coverage in future defense strategies to increase the trustworthiness of neural networks.
In Section \ref{Results}, we sum up the most important result values to demonstrate this fact.
Nonetheless, some aspects regarding detection performance, transferability, and real-world applications require further discussion. 

We have shown the mean \textit{false positive} and \textit{false negative} rates in \ref{main_results} and added the individual values to Appendix \ref{appendix_c}.
With the given values we can report a lower \textit{false negative} error throughout our proof-of-concept.
This corresponds to rather classifying benign images as adversarial than missing an attack on the target model.
Hence, we can recommend using our approach in security-sensitive setups.

Another aspect of our research worth supplemental assessment are the results during the cross-testing of various alarm model versions.
We were able to detect adversarial examples with alarm models trained using other attack methods.
With this, we give evidence for the special behaviour of neural networks when confronted with adversarial examples.
Hence, we can report two additional contributions and application scenarios.
First, future, yet unknown attacks seem detectable using our concept.
We will investigate this in future work including potentially new attacks and those which are already published.
Second, with the results gained during our cross-testing we are able to provide a ranking of the attack methods.
This ranking is based on two findings. First, the difficulty in detecting the attack, expressed by the f1-score of the respective alarm model, provides a score for rating the attack.
A second attack ranking is expressed by the performance of the alarm model created for the attack itself when detecting examples crafted with other attack methods.
As an example, let's focus on the C$\&$W attack on the ResNet target model which classifies CIFAR10 images.
We can clearly see that this attack method can only be detected by the alarm model especially created for this purpose.
Hence, we deduce the C$\&$W attack being the most powerful method.
This interpretation of our results correlates with current findings in the field of adversarial attacks and defense strategies: 
as discussed in Section \ref{related_work}, the C$\&$W attack is currently considered to be one of the most powerful attacks.

Regarding real-world application of our concept and the transferability of the approach, some aspects are worth mentioning.
In a real-world scenario DL models are often confronted with noisy inputs.
We have shown that such perturbations do not dramatically decrease the performance of the individual alarm models.
Related to this is the question about the transferability.
To give first evidence on this assumption, we tested detectability of adversarial examples in the NLP and audio processing environment with positive results.
In future work we will provide a more analysis of the performed tests in these application domains.

The second possible restriction, that our approach may suffer from, is the architecture of the model to protect.
One could argue that we are not able to detect attacks if the target model does not incorporate dense layers.
This can be ruled out considering the following.
The maintainer of the target model creates a so-called \textit{substitution model} which performs the same task as the target model itself and achieves a similar accuracy.
If this \textit{substitution model} uses dense layers, we are again able to apply our concept.
The positive results during our experiments regarding transfer attacks indicate the practicality of the \textit{substitution model}.

Finally, we want to emphasize the simplicity of our approach.
During our research on related work, we noticed some defense strategies to be rather unintuitive including several sources of errors when not applied correctly.
Our method, in contrast, is easy to use and seems intuitively reasonable.
In addition, our method does not decrease the accuracy of the model to protect when tested against benign images, which is the case for some state-of-the-art defense strategies.

\section{Conclusion} \label{Conclusion}
In this paper, we introduce a general end-to-end framework to detect adversarial examples during classification time.
Our approach consists of two main parts.

In the first phase we initialize our alarm model to detect benign and adversarial inputs during unsecured target model runtime.
In this phase, we extract neuron coverage of the target model in order to train a secondary alarm neural network. This approach is motivated by the initial hypothesis that the dense layers of the target model carry security sensitive information. 
Thus, the alarm network is trained to detect malicious activity patterns in the neurons of the target network during classification tasks.

In the second phase, the target model runs in secure operation mode, which is enabled by enhancing it with our trained secondary alarm network. 
When the target model classifies new, unseen inputs, the alarm network runs in parallel and throws an alarm if an adversarial example is being processed by the target. 
This approach leaves all parameters - especially the accuracy - of the target model untouched, yet improving overall application robustness significantly.
In our proof-of-concept implementation, we show the extensive capability of our approach to detect adversarial examples in image, NLP, and audio datasets. 
The evaluation results strongly indicate that we can not only defend with high accuracy against state-of-the-art adversarial examples, but most likely also against future, yet unknown attacks.
Finally, with adaptive attacks we show that an attacker needs to perform significantly more adversarial perturbations to successfully attack our detection-enhanced target model, compared to attacking the unsecured target model.


\bibliographystyle{plain}
\bibliography{mybib}{}

\clearpage
\appendix 
\section*{Appendix}
\section{Alarm Model Architecture} \label{appendix_a}
In the following we provide more information on the architecture of the target model we used throughout this paper.
As we said before, the alarm model is a seven-layer neural network.
The input flatten-layer accepts the concatenated extracted features, while the output layer contains two softmax-neurons in order to perform a binary classification.
As hidden layers we exclusively chose dense layers with the following amount of Relu-neurons for each layer:  \num{112}, \num{100}, \num{300}, \num{200}, \num{77}.
We trained this model for ten epochs and a batch size of \num{100}.

\section{All Result Values} \label{appendix_b}
In the following four Tables \ref{all_results_mnist_LeNet}, \ref{all_results_mnist_kerasEx}, \ref{all_results_cifar10_kerasEx}, and \ref{all_results_cifar10_resnet} we present all result values gained during the proof-of-concept.
The gray cells in each table show the accuracy and f1-score of one specific alarm model when tested against its dedicated attack method.
We emphasized the best test result in each table.

\begin{table*}[h]
\caption{All result values for the MNIST dataset and target model \textit{LeNet}.}
\begin{tabular}{|l|p{.65cm}p{.65cm}|p{.65cm}p{.65cm}|p{.65cm}p{.65cm}|p{.65cm}p{.65cm}|p{.65cm}p{.65cm}|p{.65cm}p{.65cm}|p{.65cm}p{.65cm}|}
\hline
\multicolumn{1}{|c|}{\textbf{}}                                                & \multicolumn{14}{c|}{\textbf{Accuracy and f1-Scores of the Alarm Models when tested against: (acc; f1-score)}}                                                                                                                                                                                                                                                                                                                                       \\ \cline{2-15} 
\textbf{\begin{tabular}[c]{@{}l@{}}Alarm Models trained \\ with:\end{tabular}} & \multicolumn{2}{c|}{\textit{\textbf{FGSM}}} & \multicolumn{2}{c|}{\textit{\textbf{C\&W}}} & \multicolumn{2}{c|}{\textit{\textbf{DeepFool}}} & \multicolumn{2}{c|}{\textit{\textbf{PGD}}} & \multicolumn{2}{c|}{\textit{\textbf{BIM}}} & \multicolumn{2}{c|}{\textit{\textbf{\begin{tabular}[c]{@{}c@{}}all attacks\\ combined\end{tabular}}}} & \multicolumn{2}{c|}{\textit{\textbf{\begin{tabular}[c]{@{}c@{}}transferred \\ examples\end{tabular}}}} \\ \hline
FGSM                                                                           & \cellcolor{lightgray}0.988                 & \cellcolor{lightgray}0.988               & 0.802               & 0.758               & 0.953                  & 0.945                  & 0.987                & 0.987               & 0.987                & 0.987               & 0.917                                             & 0.947                                             & 0.924                                              & 0.839                                             \\
C\&W                                                                             & 0.943                 & 0.942               & \cellcolor{lightgray}0.977               & \cellcolor{lightgray}0.977               & 0.943                  & 0.933                  & 0.957                & 0.956               & 0.958                & 0.957               & 0.944                                             & 0.965                                             & 0.951                                              & 0.905                                             \\
DeepFool                                                                       & 0.987                 & 0.987               & 0.862               & 0.843               & \cellcolor{lightgray}0.983                  & \cellcolor{lightgray}0.981                  & 0.987                & 0.987               & 0.987                & 0.987               & 0.949                                             & 0.968                                             & 0.930                                              & 0.857                                             \\
PGD                                                                            & 0.989                 & 0.989               & 0.719               & 0.615               & 0.935                  & 0.921                  & \cellcolor{lightgray}0.991                & \cellcolor{lightgray}0.991               & 0.991                & 0.991               & 0.881                                             & 0.923                                             & 0.895                                              & 0.760                                             \\
BIM                                                                            & 0.986                 & 0.986               & 0.740               & 0.655               & 0.933                  & 0.920                  & 0.990                & 0.990               & \cellcolor{lightgray}0.990                & \cellcolor{lightgray}0.990               & 0.887                                             & 0.927                                             & 0.899                                              & 0.772                                             \\
all attacks combined                                                           & 0.981                 & 0.981               & 0.973               & 0.973               & 0.977                  & 0.974                  & 0.981                & 0.981               & 0.981                & 0.981               & \cellcolor{lightgray}0.989                                             & \cellcolor{lightgray}\textbf{0.993}                                             & 0.963                                              & 0.931                                             \\
transferred examples                                                           & 0.977                 & 0.977               & 0.925               & 0.921               & 0.949                  & 0.941                  & 0.968                & 0.968               & 0.966                & 0.966               & 0.951                                             & 0.970                                             & \cellcolor{lightgray}0.961                                              & \cellcolor{lightgray}0.926                                             \\ \hline
\end{tabular}
\label{all_results_mnist_LeNet}
\end{table*}

\begin{table*}[h]
\caption{All result values for the MNIST dataset and target model \textit{kerasExM}.}
\begin{tabular}{|l|p{.65cm}p{.65cm}|p{.65cm}p{.65cm}|p{.65cm}p{.65cm}|p{.65cm}p{.65cm}|p{.65cm}p{.65cm}|p{.65cm}p{.65cm}|p{.65cm}p{.65cm}|}
\hline
\multicolumn{1}{|c|}{\textbf{}}                                                & \multicolumn{14}{c|}{\textbf{Accuracy and f1-Scores of the Alarm Models when tested against: (acc; f1-score)}}                                                                                                                                                                                                                                                                                                                                       \\ \cline{2-15} 
\textbf{\begin{tabular}[c]{@{}l@{}}Alarm Models trained \\ with:\end{tabular}} & \multicolumn{2}{c|}{\textit{\textbf{FGSM}}} & \multicolumn{2}{c|}{\textit{\textbf{C\&W}}} & \multicolumn{2}{c|}{\textit{\textbf{DeepFool}}} & \multicolumn{2}{c|}{\textit{\textbf{PGD}}} & \multicolumn{2}{c|}{\textit{\textbf{BIM}}} & \multicolumn{2}{c|}{\textit{\textbf{\begin{tabular}[c]{@{}c@{}}all attacks\\ combined\end{tabular}}}} & \multicolumn{2}{c|}{\textit{\textbf{\begin{tabular}[c]{@{}c@{}}transferred \\ examples\end{tabular}}}} \\ \hline
FGSM     & \cellcolor{lightgray}0.992       & \cellcolor{lightgray}0.992        & 0.767               & 0.700               & 0.917                  & 0.892                  & 0.992                & 0.992               & 0.985                & 0.985               & 0.890                                             & 0.929                                             & 0.893                                              & 0.792                                             \\
C\&W                                                                             & 0.973                 & 0.973               & \cellcolor{lightgray}0.975               & \cellcolor{lightgray}0.975               & 0.968                  & 0.962                  & 0.968                & 0.968               & 0.961                & 0.960               & 0.969                                             & 0.981                                             & 0.959                                              & 0.933                                             \\
DeepFool                                                                       & 0.986                 & 0.986               & 0.868               & 0.850               & \cellcolor{lightgray}0.985                  & \cellcolor{lightgray}0.982                  & 0.988                & 0.988               & 0.980                & 0.980               & 0.946                                             & 0.966                                             & 0.928                                              & 0.871                                             \\
PGD                                                                            & 0.992                 & 0.991               & 0.734               & 0.641               & 0.899                  & 0.865                  & \cellcolor{lightgray}0.992                & \cellcolor{lightgray}\textbf{0.993}               & 0.986                & 0.986               & 0.873                                             & 0.917                                             & 0.885                                              & 0.773                                             \\
BIM                                                                            & 0.992                 & 0.992               & 0.752               & 0.674               & 0.912                  & 0.886                  & 0.992                & \textbf{0.993}               & \cellcolor{lightgray}0.991                & \cellcolor{lightgray}0.991               & 0.886                                             & 0.926                                             & 0.893                                              & 0.792                                             \\
all attacks combined                                                           & 0.984                 & 0.984               & 0.972               & 0.972               & 0.980                  & 0.977                  & 0.984                & 0.984               & 0.984                & 0.984               & \cellcolor{lightgray}0.988                                             & \cellcolor{lightgray}\textbf{0.993}                                             & 0.967                                              & 0.947                                             \\
transferred examples                                                           & 0.983                 & 0.983               & 0.916               & 0.910               & 0.967                  & 0.960                  & 0.982                & 0.982               & 0.975                & 0.975               & 0.955                                             & 0.972                                             & \cellcolor{lightgray}0.966                                              & \cellcolor{lightgray}0.943                                             \\ \hline
\end{tabular}
\label{all_results_mnist_kerasEx}
\end{table*}

\begin{table*}[h]
\caption{All result values for the CIFAR10 dataset and target model \textit{kerasExC}.}
\begin{tabular}{|l|p{.65cm}p{.65cm}|p{.65cm}p{.65cm}|p{.65cm}p{.65cm}|p{.65cm}p{.65cm}|p{.65cm}p{.65cm}|p{.65cm}p{.65cm}|p{.65cm}p{.65cm}|}
\hline
\multicolumn{1}{|c|}{\textbf{}}                                                & \multicolumn{14}{c|}{\textbf{Accuracy and f1-Scores of the Alarm Models when tested against: (acc; f1-score)}}                                                                                                                                                                                                                                                                                                                                       \\ \cline{2-15} 
\textbf{\begin{tabular}[c]{@{}l@{}}Alarm Models trained \\ with:\end{tabular}} & \multicolumn{2}{c|}{\textit{\textbf{FGSM}}} & \multicolumn{2}{c|}{\textit{\textbf{C\&W}}} & \multicolumn{2}{c|}{\textit{\textbf{DeepFool}}} & \multicolumn{2}{c|}{\textit{\textbf{PGD}}} & \multicolumn{2}{c|}{\textit{\textbf{BIM}}} & \multicolumn{2}{c|}{\textit{\textbf{\begin{tabular}[c]{@{}c@{}}all attacks\\ combined\end{tabular}}}} & \multicolumn{2}{c|}{\textit{\textbf{\begin{tabular}[c]{@{}c@{}}transferred \\ examples\end{tabular}}}} \\ \hline
FGSM                                                                           & \cellcolor{lightgray}0.843                 & \cellcolor{lightgray}0.847               & 0.589               & 0.470               & 0.839                  & 0.842                  & 0.840                & 0.844               & 0.839                & 0.843               & 0.774                                             & 0.849                                             & 0.639                                              & 0.578                                             \\
C\&W                                                                             & 0.699                 & 0.679               & \cellcolor{lightgray}0.739               & \cellcolor{lightgray}0.733               & 0.662                  & 0.625                  & 0.673                & 0.642               & 0.665                & 0.631               & 0.640                                             & 0.740                                             & 0.549                                              & 0.449                                             \\
DeepFool                                                                       & 0.851                 & 0.852               & 0.571               & 0.414               & \cellcolor{lightgray}0.853                  & \cellcolor{lightgray}0.855                  & 0.853                & 0.855               & 0.853                & 0.855               & 0.767                                             & 0.843                                             & 0.624                                              & 0.540                                             \\
PGD                                                                            & 0.839                 & 0.840               & 0.584               & 0.449               & 0.841                  & 0.842                  &\cellcolor{lightgray}0.841                &\cellcolor{lightgray}0.843               & 0.841                & 0.843               & 0.762                                             & 0.840                                             & 0.623                                              & 0.543                                             \\
BIM                                                                            & 0.848                 & 0.850               & 0.580               & 0.434               & 0.849                  & 0.850                  & 0.849                & 0.851               & \cellcolor{lightgray}0.850                & \cellcolor{lightgray}0.852               & 0.767                                             & 0.844                                             & 0.618                                              & 0.530                                             \\
all attacks combined     & 0.708    & 0.771               & 0.678               & 0.741               & 0.709                  & 0.771                  & 0.709                & 0.772               & 0.709                & 0.772               & \cellcolor{lightgray}0.882       & \cellcolor{lightgray}\textbf{0.932}            & 0.688                                              & 0.754                                             \\
transferred examples                                                           & 0.749                 & 0.777               & 0.577               & 0.557               & 0.738                  & 0.766                  & 0.743                & 0.771               & 0.743                & 0.770               & 0.769                                             & 0.852                                             & \cellcolor{lightgray}0.704                                              & \cellcolor{lightgray}0.732                                             \\ \hline
\end{tabular}
\label{all_results_cifar10_kerasEx}
\end{table*}

\begin{table*}[h]
\caption{All result values for the CIFAR10 dataset and target model \textit{ResNet}.}
\begin{tabular}{|l|p{.65cm}p{.65cm}|p{.65cm}p{.65cm}|p{.65cm}p{.65cm}|p{.65cm}p{.65cm}|p{.65cm}p{.65cm}|p{.65cm}p{.65cm}|p{.65cm}p{.65cm}|}
\hline
\multicolumn{1}{|c|}{\textbf{}}                                                & \multicolumn{14}{c|}{\textbf{Accuracy and f1-Scores of the Alarm Models when tested against: (acc; f1-score)}}                                                                                                                                                                                                                                                                                                                                       \\ \cline{2-15} 
\textbf{\begin{tabular}[c]{@{}l@{}}Alarm Models trained \\ with:\end{tabular}} & \multicolumn{2}{c|}{\textit{\textbf{FGSM}}} & \multicolumn{2}{c|}{\textit{\textbf{C\&W}}} & \multicolumn{2}{c|}{\textit{\textbf{DeepFool}}} & \multicolumn{2}{c|}{\textit{\textbf{PGD}}} & \multicolumn{2}{c|}{\textit{\textbf{BIM}}} & \multicolumn{2}{c|}{\textit{\textbf{\begin{tabular}[c]{@{}c@{}}all attacks\\ combined\end{tabular}}}} & \multicolumn{2}{c|}{\textit{\textbf{\begin{tabular}[c]{@{}c@{}}transferred \\ examples\end{tabular}}}} \\ \hline
FGSM & \cellcolor{lightgray}0.810  & \cellcolor{lightgray}0.815     & 0.559  & 0.446  &  0.812 & 0.820  &  0.812 & 0.821  &  0.812 & 0.821  &  0.761  & 0.809    & 0.658  & 0.674     \\
C\&W   & 0.701 & 0.715  & \cellcolor{lightgray}0.707 & \cellcolor{lightgray}0.727  & 0.698  & 0.716  & 0.697 & 0.716   & 0.695 & 0.713 & 0.725 & 0.789 & 0.703 & 0.753          \\
DeepFool & 0.819   & 0.827   & 0.568  & 0.469  & \cellcolor{lightgray}0.822   &\cellcolor{lightgray}0.833  & 0.823    & 0.835 & 0.823  & 0.834  & 0.778 & 0.826 & 0.666    & 0.686        \\
PGD     & 0.822    & 0.826     & 0.560  & 0.434 & 0.824   & 0.831  & \cellcolor{lightgray}0.826  & \cellcolor{lightgray}0.833 & 0.825  & 0.832  & 0.767  & 0.813   & 0.644   & 0.651    \\
BIM  & 0.819  & 0.825    & 0.559  & 0.445  & 0.821   & 0.830 & 0.822  & 0.832   & \cellcolor{lightgray}0.822     & \cellcolor{lightgray}0.832    & 0.771   & 0.818  & 0.646    & 0.658         \\
all attacks combined & 0.788  & 0.812    & 0.677     & 0.687    & 0.792     & 0.818     & 0.793       & 0.820      & 0.793      & 0.819    & \cellcolor{lightgray}0.812       & \cellcolor{lightgray}\textbf{0.864}    & 0.744      & 0.791               \\
transferred examples  & 0.672 & 0.614   & 0.570  & 0.440  & 0.655 & 0.593  & 0.655 & 0.595 & 0.653  & 0.592 & 0.583      & 0.602     & \cellcolor{lightgray}0.662  & \cellcolor{lightgray}0.670       \\ \hline
\end{tabular}
\label{all_results_cifar10_resnet}
\end{table*}

\section{Confusion Matrix Values} \label{appendix_c}
In Table \ref{all_conf_matrices} we show the confusion matrix values of each performed test.
\begin{table*}[h]
\centering
\caption{Confusion Matrix values for all datasets, target models, and attack methods. Each result belongs to the detection of adversarial attack methods with the according target model.}
\begin{tabular}{|l|l|l|l|l|l|l|}
\hline
\multirow{2}{*}{\textbf{Dataset}} & \multirow{2}{*}{\textbf{Target Model}} & \multirow{2}{*}{\textbf{Attack}} & \multicolumn{4}{p{7cm}|}{\textbf{Performance of the Alarm Models when tested against the according attack method}}                         \\ \cline{4-7} 
                                  &                                             &                                       & \textit{\textbf{True Positive}} & \textit{\textbf{True Negative}} & \textit{\textbf{False Positive}} & \textit{\textbf{False Negative}} \\ \hline
MNIST                             & LeNet                                       & FGSM                                  & 1.00                               & 0.98                            & 0.02                             & 0.00                                \\
                                  &                                             & C\&W                                    & 0.98                            & 0.97                            & 0.03                             & 0.02                             \\
                                  &                                             & DeepFool                              & 0.99                            & 0.98                            & 0.02                             & 0.01                             \\
                                  &                                             & PGD                                   & 0.99                            & 0.99                            & 0.01                             & 0.01                             \\
                                  &                                             & BIM                                   & 0.99                            & 0.99                            & 0.01                             & 0.01                             \\ \cline{2-7} 
                                  & kerasExM                                    & FGSM                                  & 1.00                               & 0.99                            & 0.01                             & 0.00                                \\
                                  &                                             & C\&W                                    & 0.98                            & 0.97                            & 0.03                             & 0.02                             \\
                                  &                                             & DeepFool                              & 0.99                            & 0.98                            & 0.02                             & 0.01                             \\
                                  &                                             & PGD                                   & 1.00                               & 0.99                            & 0.01                             & 0.00                                \\
                                  &                                             & BIM                                   & 0.99                            & 0.99                            & 0.01                             & 0.01                             \\ \cline{2-7} 
                                  & LSTM                                        & Transfer                              &  0.97                           & 0.88                            & 0.12                       & 0.03                                 \\ \cline{2-7} 
                                  & CapsuleNN                                   & Transfer                              &  1.00                               & 1.00                             &   0.00                               & 0.00                                 \\ \hline
CIFAR10                           & kerasExC                                    & FGSM                                  & 0.87                            & 0.81                            & 0.19                             & 0.13                             \\
                                  &                                             & C\&W                                    & 0.72                            & 0.76                            & 0.24                             & 0.28                             \\
                                  &                                             & DeepFool                              & 0.87                            & 0.84                            & 0.16                             & 0.13                             \\
                                  &                                             & PGD                                   & 0.85                            & 0.83                            & 0.17                             & 0.15                             \\
                                  &                                             & BIM                                   & 0.86                            & 0.84                            & 0.16                             & 0.14                             \\ \cline{2-7} 
                                  & ResNet                                      & FGSM                                  & 0.86                            & 0.76                            & 0.24                             & 0.14                             \\
                                  &                                             & C\&W                                    & 0.78                            & 0.63                            & 0.37                             & 0.22                             \\
                                  &                                             & DeepFool                              & 0.89                            & 0.75                            & 0.25                             & 0.11                             \\
                                  &                                             & PGD                                   & 0.87                            & 0.78                            & 0.22                             & 0.13                             \\
                                  &                                             & BIM                                   & 0.88                            & 0.76                            & 0.24                             & 0.12                             \\ \hline
\end{tabular}
\label{all_conf_matrices}
\end{table*}

\section{Adversarial Example Generation for the NLP scenario} \label{appendix_d}
With Algorithm \ref{alg_for_nlp_adv} we generated adversarial examples for our target model classifying IMBD reviews.
The target model performs a binary classification and tries to distinguish between positive and negative reviews respectively.

\begin{algorithm}[]
    \KwData{IMDB reviews}
    \KwResult{adversarial IMDB reviews}
    train a Word2Vec Model with all reviews\;  
    randomly pick one word to start\;
    \While{not at the end of this document}{
        find $N$ most similar words of current word with Word2Vec\;
        \For{substitute $\leftarrow$ \textit{next most similar word}}{
            replace the current word with the \textit{substitute}\;
            predict and calculate the margin\;
            \If{margin decrease}{
                \textbf{break}
            }
        }
        \eIf{margin $<$ \textit{MarginThreshold}}{
            \textbf{break}
        }{
            recover the current word to the original word\;
            move to next word\;
        }
    }
    \caption{Generation of adversarial examples in the IMDB dataset containing movie reviews.}
    \label{alg_for_nlp_adv}
\end{algorithm}
\clearpage
\section{C\&W Attack Parameters for the Adaptive Attack} \label{appendix_e}
Table \ref{adaptive_attack_params_mnist} shows the attack parameters of the C\&W attack during the adaptive white-box attacks for the MNIST dataset.
\begin{table}[!h]
\begin{tabular}{|l|l|}
\hline
Parameter Name      & Parameter Value \\ \hline
max-iteration       & \num{3000}      \\
batch-size          & \num{100}       \\
learning-rate       & \num{0.005}     \\
binary-search-steps & \num{20}        \\ \hline
\end{tabular} 
\caption{C\&W attack parameters during the adaptive attack for MNIST.}
\label{adaptive_attack_params_mnist}
\end{table}

Table \ref{adaptive_attack_params_cifar} shows the attack parameters of the C\&W attack during the adaptive white-box attacks for the CIFAR10 dataset.
\begin{table}[h]
\begin{tabular}{|l|l|}
\hline
Parameter Name      & Parameter Value \\ \hline
max-iteration       & \num{100}      \\
batch-size          & \num{100}       \\
learning-rate       & \num{0.01}     \\
binary-search-steps & \num{5}        \\ \hline
\end{tabular} 
\caption{C\&W attack parameters during the adaptive attack for CIFAR10.}
\label{adaptive_attack_params_cifar}
\end{table}

\end{document}